\documentstyle[preprint,aps,epsf]{revtex}

\newcommand{\referencestyle}{
\small
\abovedisplayskip=6pt
\belowdisplayskip=6pt
\vspace{12pt}}

\def\De{\Delta\eta}
\def\D{\Delta}
\def\Det{\Delta\eta_T}

\def\Dy{\Delta y}
\def\t0{\tau_0}

\def\ch{\cosh}
\def\sh{\sinh}
\def\ben{\begin{eqnarray}}
\def\enn{\end{eqnarray}}
\def\ov{\over\displaystyle\strut}
\def\l({\left(}
\def\r){\right)}
\def\dst{\displaystyle\strut}
\def\lpt{~low-$p_t$~}
\def\lpte{LPTE}
\def\en{enhancement~}

\begin{document}

\rightline{LUNFD6/(NFFL-7092) 1994}
\rightline{hep-ph/9412323}


\begin{center}
{\large \bf Rapidity-Dependent Low-$p_t$ Enhancement
}
\end{center}

\medskip

\begin{center}
{\rm T. Cs\"org\H o$^{1,2}$
}
\end{center}

\begin{center}
{\it
$^1$MTA KFKI RMKI \\
H--1525 Budapest 114, P.O. Box 49. Hungary \\
$^{2}$Department of Elementary Particle Physics,  \\
University of Lund,
S\"olvegatan 14,
S - 223 62 Lund, Sweden
}
\end{center}
\begin{center}
\today
\end{center}

\begin{abstract}
The rapidity dependence of the low-$p_t$ enhancement
is shown to be a sensitive measure of the
longitudinal source size for longitudinally expanding finite systems.
\end{abstract}


Low-$p_t$ \en (LPTE)
is conventionally understood as a rise in the
ratio of the transverse momentum distribution of the particles produced
in proton-nucleus ($p + A$) as well as
heavy ion ($A + B$) reactions
as compared to the transverse momentum distribution
of the same particles in hadron-hadron reactions at the same
center of mass energy per bombarding particle.
An up-to-date summary of the status of the experimental data
and theoretical understanding of LPTE can be
found in ref.~\cite{gillo}
 which concluded: " In order to understand
 the results, theoretical models should ... explain the
 rapidity dependence observed in the data."

In a recent paper
the invariant momentum distribution (IMD) and the
Bose-Einstein correlation function (BECF) has been calculated
for longitudinally expanding finite systems~\cite{1d}.
As an application of those results it is
shown here that  the longitudinal expansion
together with the finite longitudinal size of
the expanding tube naturally leads to a {\it rapidity-dependent}
LPTE.

Resonance decay effects~\cite{reso1,reso2} and other, more
exotic  sources of LPTE,
like the decay of droplets of Quark-Gluon Plasma~\cite{drops},
or the effects of attractive potentials~\cite{pots} leading to a change in the
dispersion relation shall
modify primarily the transverse momentum distribution.
However, these modifications are due to
a mechanism which is independent (modulo isospin) of the
longitudinal expansion thus the rapidity dependence
of the LPTE does not naturally follows
from these models.
Partial thermal equilibrium due to a longitudinal expansion
was shown to lead to additional soft pions in ref.~\cite{long}.
These models explain neither the point why the enhancement
is similar in $p + A$ and $A+ A$ collisions
nor the observed rapidity dependence of the \lpte ~data.

Instead of studying the $p_t$ spectrum directly, let us
 concentrate on the coupling between the
transverse momentum spectra and the {\it longitudinal}
expansion.

\medskip

{\it Longitudinally expanding finite systems}.
High energy heavy ion collisions
may create one-dimensionally
expanding systems, especially in the case of light projectiles.
Heavier projectiles may create three-dimensionally
expanding systems. However, the three-dimensional
expansion does not alter the coupling between
transverse momentum spectra and rapidity
distribution in the limit $m_t \rightarrow m$
as long as the expansion remains in the same limit
longitudinally boost-invariant.

The four-momentum is given by $p = (E, \vec p\,) = (E, p_x, p_y, p_z)$.
The particle is on the mass shell $ m^2 = E^2 - \vec p^{\,\, 2} $,
the transverse mass is denoted by $m_t = \sqrt{E^2 - p_z^2}$,
 the rapidity by $y = 0.5 \ln\l({\displaystyle\strut
E + p_z \ov E - p_z}\r)$, the space-time rapidity by
$\eta =  0.5 \ln\l({\displaystyle\strut t + r_z \ov t - r_z} \r)$
and the transverse momentum is indicated by $ p_t = \sqrt{ p_x^2 + p_y^2}$.

If a locally thermalized relativistic momentum
distribution creates the correlations
between rapidity and space-time
rapidity, the one-particle IMD can be calculated
for arbitrary space-time distribution functions as
\ben
{\displaystyle\strut d^2n\ov dy dm_t^2} & = &
        g(m_t) \, \, \int_{-\infty}^{\infty} d\eta \,\,
         G(\eta) \, J_{m_t}(\eta, y).
        \label{e:dng}
\enn
In this equation the transverse and temporal components of the
space-time distribution have already been integrated over~\cite{1d},
and a finite size in the space-time rapidity is introduced by the
distribution function $G(\eta)$. The various rapidity-independent
modifications of the low-$p_t$ spectrum are taken into account
through the factor $g(m_t)$ which shall be discussed later. Moreover,
the  local relativistic Bose-Einstein or Fermi-Dirac distributions
for longitudinally expanding systems are given by
\ben
J_{m_t}(\eta, y) \, d\eta & = & {\dst f  \ov (2 \pi)^3 }
{\dst d\sigma(\eta) \cdot p \ov
	\exp\l(  p \cdot u(\eta) / T \r) \pm 1}.
	\label{e:befd}
\enn
Here $f$ is the degeneracy factor,
$ d\sigma(\eta) \cdot p$ is the inner-product of the volume-element
of the freeze-out hypersurface and the four-momentum $p = (\, m_t \cosh(y),
\, p_x,\, p_y,\, m_t \sinh(y) \,)$, while
$u(\eta)$ stands for the four-velocity of the expanding matter
on the freeze-out hypersurface and $T$ is the freeze-out temperature.
The $+$ sign stands for fermions, the $-$ sign for bosons.

For the four-velocity of the 1D expanding matter
we assume the scaling Bjorken profile~\cite{bjorken} as
\ben
u(\eta) & = & (\ch (\eta),\, 0,\, 0,\, \sh(\eta) ). \label{e:u}
\enn
Accordingly, $J_{m_t}(\eta,y)$
becomes a function of $\eta - y$ only.

All the conventional and non-conventional modifications
of the $m_t$ distribution which are independent of
the rapidity and aim at describing the low-$p_t$ behaviour at mid-rapidity,
enter into the function $g(m_t)$.
This function has influence on the
rapidity dependence of the effective temperature.
We assume that
\ben
      g(m_t) &  = & C \exp( - m_t / T_* )
\enn
where $T_*$ stands for a {\it rapidity-independent} effective
temperature in the \lpt limit.
Such a structure
is the simplest limiting distribution of many complex
calculations attempting to describe the LPTE.
Note that here we are interested only in the \lpt behaviour and
thus a deviation of the $g(m_t)$ distribution from the
above simple exponential for medium or high values of
$m_t$ is not relevant for our considerations.
Thus, the phenomenological parameter $T_*$ stands for
all the possible rapidity independent modifications of the
$m_t$ distribution at low-$p_t$, including decays of resonances,
possible modifications of the dispersion relations and possible
more exotic effects.

In this model, the source at a given space-time -- rapidity $\eta $
is emitting bosons with the mean rapidity $ \langle y \rangle =\eta$.
 The width of the emission can be calculated by
approximating the Bose-Einstein and the Fermi-Dirac
distributions with the relativistic Boltzmann-distribution,
\ben
 J_{m_t}(\eta,y)
  &  \approx  &
        C'(m_t)\, \exp\l( - m_t (y - \eta)^2 /T \r)
                \label{e:b}
\enn
where $C'(m_t) \equiv \, C \, m_t \, \exp(-m_t/T)$ is independent of
rapidity and
can thus be absorbed into the $g(m_t)$ factor.
Only the leading order
$m_t (\eta - y)^2 /T$ term has been kept
in the Taylor expansion of the logarithm of $J_{m_t}$.
Terms of ${\cal O}\l((\eta - y)^2\r)$
 are next to leading order since the Boltzmann approximation
is valid for $m_t / T  >> 1$ only.
 A numerical comparison of the approximated
to its approximation indicates that the approximation is excellent
for $m_t > 3 T$ and reasonable down to
$m_t > 1.2 T$.
For  certain classes of the models~\cite{nr,3d}
the freeze-out temperature $T$ is smaller than the
effective temperature at mid-rapidity,
$T_* = T + T_G$. Here the geometrical contribution to the
effective temperature is given in the low-$p_t$ limit by
$T_G = m R_G^2 a^2 / \t0^2$ where
 the finite transverse geometrical size is $R_G$,
 the mean freeze-out time
is denoted by
$\tau_0$ and $a$ stands for the flow gradinent in units of the mean freeze-out
time.
A preliminary analysis of the pion IMD
{}~\cite{qm95} indicates that
$m_{\pi}/T \approx 1.5$, for kaons $m_{K} / T \approx 6$,
thus the approximation ~(\ref{e:b}) is warranted.

A   Gaussian approximation is also applied
for the distribution function of $\eta$
\ben
        G(\eta) & =  & {\displaystyle\strut 1\ov (2 \pi \De^2)^{1/2} }\,
       \exp\l( - {\displaystyle\strut (\eta - y_0)^2 \ov 2 \De^2} \r) ,
		\label{e:g}\\
        J_{m_t}(\eta,y) & = &
           {\displaystyle\strut 1\ov (2 \pi \Det^2(m_t))^{1/2} }\,
        \exp\l(\displaystyle\strut - {(\eta - y)^2 \ov 2 \Det^2(m_t)} \r),
                \label{e:det}
\enn
where
\ben
\Det\,(m_t) & = & \sqrt{T / m_t},
\enn
the mid-rapidity is denoted by $y_0$
and the width of the space-time rapidity
distribution is given by $\De$,  which is a dimensionless
measure of the longitudinal extension
of the expanding system at freeze-out.
By introducing this finite width $\De$, we break
the boost-invariance of our source in the longitudinal direction too.
Thus we expect a non-stationary rapidity
distribution.  The flat rapidity
distribution corresponds to
the $\De \rightarrow \infty $ limit.

{\it Gaussian results}.
The IMD  can be calculated~\cite{1d} as
\ben
{\displaystyle\strut d^2 n\ov dy dm_t^2 }
  & = & {\displaystyle\strut 1 \ov (2 \pi \Dy^2(m_t)\,)^{1/2}}
		  \,\, g(m_t)\,\,
   \exp( - {\displaystyle\strut (y-y_0)^2 \ov 2 \Dy^2(m_t)}),
	\label{e:imd}
\enn
where
\ben
        \Dy^2(m_t) & = & \De^2 + \Det^2(m_t).
\enn
	This relation introduces
	a coupling between the variables $y$ and $m_t$,
	i.e. their distributions do not factorize any more.
	At a fixed value of the rapidity, let us expand eq.~(\ref{e:imd})
	around $m_t = m$ as
\ben
	{\dst d^2n \ov dy \, dm_t^2} & \propto &
		\exp\l( - {\dst m_t \ov  T_e(y)} +
	{\cal O}\l(
	(m_t - m)^2 / m^2\r) \r)
\enn
which yields  a Lorenzian effective temperature distribution
in the \lpt region:
\ben
	T_e(y) & = & {\dst T_* \ov 1 + a (y - y_0)^2},
		\label{e:ty}\\
	a & = & {\dst T_* T \ov 2 m^2}
		\l( \De^2 + {\dst T \ov m} \r)^{-2}, \label{e:aa}
\enn
a dependence which is similar to the recent finding
of the NA35 collaboration for the charged hadrons in $S + Pb$
collisions at CERN SPS~\cite{mitchell}.
Such a change in the effective temperature with
rapidity is a general property of heavy ion reactions
performed both
at CERN SPS and at the AGS, as measured by the NA35,
E802/E859, E810, E814 collaborations valid
pions, kaons protons and lambdas, see Fig. 3. in ref.~\cite{stachel}.
Thus the decrease of $T_e(y)$ in the
target and projectile fragmentation region can be considered
as a simple consequence of both the longitudinal expansion in the
final stage and the $m_t$ dependent width of the
local thermal rapidity distribution. Simple kinematics
results in a decrease of the average transverse momentum in the
target and projectile fragmentation regions for $p + p$
reactions too, see Fig. 9 of ref.~\cite{wang}.

With $T_e(y) $ given above the
IMD can be integrated over
$m_t$ at a fixed $y$, resulting
in
\ben
{\dst dn\ov dy} & = & {\cal N}\, {\dst T_e(y) \ov T_* } \,\,
			{ \dst m + T_e(y) \ov m + T_*} \,\,
		\exp\l( - {\dst (y - y_0)^2 \ov 2 \l( \De^2 +
				T / m  \r) } \r),
	\label{e:dndy}
\enn
where ${\cal N}$ is a normalization constant.
This distribution describes for small values of $a$
an approximately Gaussian rapidity distribution,
with corrections
which reduce the effective width of the
Gaussian to $\Dy^2_{eff} < \De^2 +  {\dst T \ov m } $.

Eqs.~(\ref{e:ty},\ref{e:dndy})
are given in terms
of three free parameters, such as the (dimensionless) longitudinal
size $\D\eta$, the freeze-out temperature $T$ and the effective
temperature $T_*=T_e(y_0)$. Thus the longitudinal size $\D \eta$,
which is hardly accessible to HBT measurements~\cite{1d,3d},
can be determined from momentum space measurements alone!

The experimental analysis of LPTE
is based on a comparison of  the IMD in $p+A$ or $A + B$ reactions
to the IMD in $p+p$ reaction at the same energy,
\ben
R^{AB}_{pp} & = & {\dst d^2n_{A+B} \ov dy \, dm_t^2}
		\l({\dst d^2n_{p+p} \ov dy \, dm_t^2}\r)^{-1}.
\enn
This ratio can easily be evaluated from eq.~(\ref{e:imd})
as
\ben
R^{AB}_{pp}& = &
  \exp\l(-{\dst m_t \ov T_*^{AB} }+{\dst m_t \ov T_*^{pp} } \r)
	\times
   \nonumber \\
   \null & \null &
  \exp\l( + {\dst (y - y_0)^2 (\D^2\eta^{pp} - \D^2\eta^{AB}) \ov
	( \D^2\eta^{pp} + T/m_t) \, (\D^2\eta^{AB} + T / m_t) } \r)
	\label{e:labpp}
\enn
If $T$ is
independent of the reaction type,
 $R^{AB}_{pp}$ contains two
factors. The first one is independent of the rapidity and
 may account for the various modifications of the $p_t$ spectrum
at mid-rapidity due to rescattering, nuclear resonance production etc.
This factor may yield the LPTE at mid-rapidity.
The second factor determines the rapidity dependence of the effect.
Generally, the degree of stopping in $p+p$ reactions is smaller
than in $p+A$ or $A + B$ reactions at the same energy,
due to the enhanced rescattering in the second and third case.
This also implies $ \D\eta^{pp} >  \D\eta^{pA} \geq \D\eta^{AB}$,
consequently
indicate an increase in $R^{AB}_{pp}$ within the target and projectile
fragmentation region.
Thus the LPTE in $A+B$ or $p+A$ reactions as
compared to  $p+p$ reactions becomes a measure
of the {\it difference} in the longitudinal sizes
for the two reactions.

 Can the LPTE be utilized
to measure the  {\it total} longitudinal
source sizes (not only differences) ?

Let us introduce as an
auxiliary quantity the ratio of the IMD-s
for rapidities $y_1$ and $y_2$
at the same $m_t$
{\it in the same reaction},
either $p+p$ or $p+A$ or $A+B$
\ben
R(y_1,y_2,m_t) & = & {\dst d^2 n \ov dy_1 \, dm_t^2 }
		  \l( {\dst d^2 n \ov dy_2 \, dm_t^2 } \r)^{-1} .
\enn
The rapidity-independent part of the momentum-distribution, $g(m_t)$
cancels from this ratio:
\ben
R(y_1,y_2,m_t) & = &  {\dst \int d\eta \, G(\eta) \, J_{m_t}(\eta,y_1)
			\ov  \int d\eta \, G(\eta) \, J_{m_t}(\eta,y_2) }.
\enn

Let us introduce the normalized ratio
of the invariant momentum distributions,
\ben
L_R (y, m_t) & = & {\dst R(y,y_0,m_t) \ov
		\lim_{m_t \rightarrow \infty}  R(y,y_0,m_t)}
\enn
where $y_0$ stands, as before, for the mid-rapidity.
This normalized ratio shall be a measure of the LPTE
 as compared to the \lpt distribution at mid-rapidity.
The normalized ratio $L_R $ has the property
\ben
\lim_{m_t \rightarrow \infty} L_R(y, m_t) & = &
L_R( y_0, m_t)  \equiv    1.
\enn
Although the limit $m_t \rightarrow \infty$ formally appears
in the definition of $L_R$, the
limiting value of $L_R(y,m_t)$ is
1, independently of the rapidity.
This asymptotic value can be utilized  for guiding the
experimental normalization of this quantity.
In the Gaussian approximation, $L_R$
reads as
\ben
L_R(y,m_t) & = & \exp\l( + {\dst (y - y_0)^2 \ov 2 \De^2}
		{\dst T \ov m_t \l(\De^2 +  T / m_t\r) }
				\r)
				\label{e:lg}
\enn
The LPTE becomes
 by eqs.~(\ref{e:labpp},\ref{e:lg})
{\it exponentially enhanced}
in the target and projectile fragmentation region.
This is {\it due to the decrease of $T_e(y)$ }
in the same region, cf. eq.~(\ref{e:ty}).
As a consequence, the \lpt ratio $L_R(y,m_t)$
becomes a sensitive measure of both
the geometrical size
of the emission region in the longitudinal direction
$\Delta \eta$
and the freeze-out temperature $T$.
The LPTE has been calculated without explicit
reference to the particle type, thus one might expect
LPTE to exists for pions, kaons as well as for protons and any other
heavier particles.
 At a given value of $p_t$,
 heavy particles have a substancially reduced
LPTE compared to pions
according to eq.~(\ref{e:lg}).

The expression~(\ref{e:lg}) indicates that the
rapidity-dependent LPTE
is a consequence of the finite size of the longitudinally
expanding system, because
the enhancement vanishes
in the limit the longitudinal size goes to infinity,
\ben
\lim_{\De \rightarrow \infty} L_R(y,m_t) & = & 1,
\enn
independently of $y$ and $m_t$.
Thus it is more natural
to compare the IMD, at a given rapidity,
to the IMD at mid-rapidity in the same reaction,
than to compare the IMD in $p+A$ or $A + B$ reactions
to the IMD in $p+p$ reaction, at the same energy.
However, this latter possibility has been realized in
most of the available data analysis on LPTE.
As can be seen from eq.~(\ref{e:labpp}),
the rapidity dependence of the ratio $R^{AB}_{pp}$
is also a consequence of the finite longitudinal
sizes of the expanding systems since
\ben
\lim_{ \begin{array}{l}\scriptstyle
\De^{pp}, \De^{AB} \rightarrow \infty, \\
 \scriptstyle(\De^{pp} - \De^{AB}) < \infty
       \end{array}}
	 R^{AB}_{pp}
  & = &
\exp\l(-{\dst m_t \ov T_*^{AB} }+{\dst m_t \ov T_*^{pp} } \r),
\enn
which is rapidity independent.

The results presented here are insensitive
to resonance decay effects if these are not coupled to
a specific rapidity region, since in such cases
the rapidity independent effective temperature, $T_*$
may absorb their influence. A notable exception
can be the delta resonance, which is copiously produced
in the target and projectile fragmentation region.
Since the effect of the longitudinal expansion is
increasing with $y-y_0$ like an inverse Gaussian,
eqs.~(\ref{e:lg},\ref{e:labpp}), only
a comparably strong coupling between the
$\Delta$ production and the rapidity can influence it
significantly. Such a strong coupling would be, however, rather
unusual.

The result for the LPTE is insensitive to the presence
of the transverse flow, since similar results
can be obtained for the IMD of ref.~\cite{3d}, where the flow profile
was approximated by a {\it three-dimensional} scaling
flow which describes a fully developed
transverse flow. The transverse flow cancels since
it influences the transverse momentum spectra only at  high
values of $m_t$.

It is straightforward to generalize the above results for
arbitrary $G(\eta) $ functions,
to be  published elsewhere.

{\it Application}. The E802 collaboration has found a characteristic
bell-shaped curve for $T_e(y)$
in central $ S + Au$ reactions at 14.6 AGeV ~\cite{e802}.
It should be clear from the previous parts that the decrease
of $T_e(y)$ with increasing $\mid y-y_0\mid $ results in a
rapidity-dependent LPTE. Thus   a fit to $T_e(y)$ is
a kind of measure of the rapidity-dependece ot LPTE.
The E802 data are fitted with eqs.~(\ref{e:ty},\ref{e:aa}) in Fig. 1.
The data are compatible with our description at a $\chi^2 /NDF = 1.93$.
This result is somewhat surprising since we have utilized the scaling flow
profile to describe the longitudinal expansion.
Note, however, that such a flow profile may develop not only in
reactions where the incoming nuclei pass through each other,
but also may develop in the final stage of the hydrodynamical
evolution in the stopping region~\cite{levai}.
The mid-rapidity at BNL AGS is at $y_{mid} \approx 1.7$
for symmetric reactions, the fit yields somewhat smaller value of
$y_0 = 1.52 \pm 0.25$ for the asymmetric $S + Au $ collision.
The mid-rapidity temperature is found to be $T_* = 209 \pm 10$
MeV, the freeze-out temperature cannot be determined precisely
at the present level of the experimetal errors $T = 60 \pm 418$ MeV.
Finally, we measure the length of the kaon-emitting
volume at the freeze-out time, in dimensionless
space-time rapidity units to be $\Delta \eta_K = 0.39 \pm 0.18$.

 {\it In summary}, the model predicts a rapidity-dependent low-$p_t$
	enhancement
	for $p + p$, $p+A$ and $A+B$ reactions, when
	compared to the transverse mass spectrum at mid-rapidity
	in the same reactions.
	The relative enhancement in $p + A$ and $A + B$ reactions
	compared to the transverse momentum spectrum in $p + p$
	reactions at the same energy turns out to be a
	consequence of the increased degree of stopping in $p+A$
	and $A+B$ reactions as compared to the stopping in $p + p$
	collisions.

 The rapidity dependence of the low-$p_t$
 enhancement can be used to measure the
 finite longitudinal size for longitudinally expanding systems.
 This method is especially advantageous since the large
 longitudinal geometrical sizes of expanding systems
 appear in the radius parameters
 of Bose-Einstein correlation measurements
 as correction terms only~\cite{1d,nr,3d}.
	Using a variation of this technique we have determined
	the length of the $K^+$ emitting volume at freeze-out
	in central $S + Au$ reactions at AGS to be
	$\Delta\eta_{K^+} = 0.39 \pm 0.18$ space-time rapidity
	unit.

{\it Acknowledgments:}
Thanks are due to X.-N. Wang,
 B. L\"orstad and G. Gustafson for kind hospitality and stimulation
during the author's stay at Lawrence Berkeley Laboratory and at
University of Lund and to T. Dolinszky for a careful reading of
the manuscript.
This work has been supported
by the Human Capital and Mobility (COST) programme of the
European Economic  Community under grants No. CIPA - CT - 92 - 0418
(DG 12 HSMU), by the Hungarian
NSF  under Grant  No. OTKA-F4019
and by the Hungarian - U. S. Joint Fund under grant  MAKA 378/93.

\null\vfill
\begin{figure}
          \begin{center}
          \leavevmode\epsfysize=4.5in
          \epsfbox{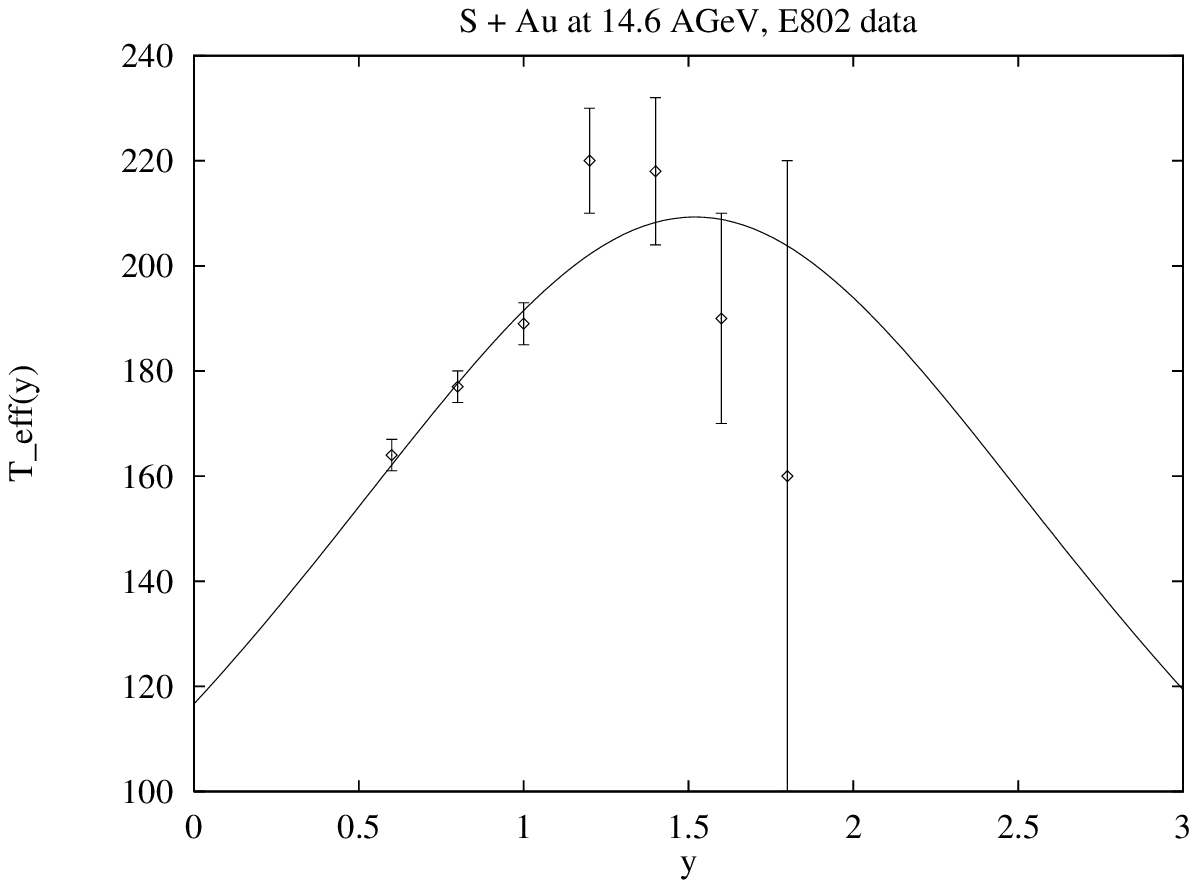}
          \end{center}
\caption{\label{f:1}}
{\small
 Fit to $T_e(y)$ of $K^+$ mesons in 14.6 AGeV $S + Au$ reactions
with eqs.~(\ref{e:ty},\ref{e:aa}),
solid line. Data points with error bars have been scanned
from ~\cite{e802}.
 The fit parameters are $\Delta\eta= 0.39 \pm 0.18$,
 $T_* = 209 \pm 10$ MeV, $T = 60 \pm 418 $ MeV and $y_0 = 1.52 \pm 0.25$
for the minimum of $\chi^2/NDF = 5.8/ (7-4) = 1.93$ using $m_K = 494$ MeV.
}
 \end{figure}
\vfill\eject

\begin{thebibliography}{99}
\referencestyle
\bibitem{gillo}         J. Simon-Gillo, Nucl.\ Phys.\ {\bf A566} (1994) 175c.
\bibitem{1d}            T. Cs\"org\H o, hep-ph/9409327,
                        LUNFD6/(NFFL-7081) 1994,
                        (Phys. Lett. B., submitted).
\bibitem{bjorken}       J. D. Bjorken, Phys. Rev. {\bf D27} (1983) 140.
\bibitem{reso1}         J. Sollfrank et al, Phys. Lett. {\bf B252} (1990) 256.
\bibitem{reso2}         G. E. Brown et al, Phys. Lett. {\bf B253} (1991) 19.
\bibitem{drops}         L. van Hove,  Ann. Phys. 192 (1989) 66.
\bibitem{pots}          E. V. Shuryak, Phys. Rev. {\bf D42} (1990) 1764.
\bibitem{long}          S. Gavin \& P. Ruuskanen,
			Phys.Lett. {\bf B262} (1991) 326.
\bibitem{nr}            T. Cs\"org\H o, B. L\"orstad and J. Zim\'anyi,
                        Phys. Lett. {\bf B338} (1994) 134 .
\bibitem{3d}            T. Cs\"org\H o and B. L\"orstad,
                        LUNFD6/(NFFL-7082) 1994,
			(Phys. Rev. Lett., submitted).
\bibitem{qm95}		T. Cs\"org\H o, Quark Matter'95 conference,
			Nucl. Phys. A (in preparation).
\bibitem{mitchell}	J. Mitchell et al, Nucl. Phys. {\bf A566} (1994) 415c.
\bibitem{stachel}       J. Stachel,  Nucl.\ Phys.\ {\bf A566}, (1994) 183c.
\bibitem{wang}		X.-N. Wang \& M. Gyulassy, Phys.Rev. {\bf D44} (1991)
			3501.
\bibitem{e802}		D. P. Morrison et al,Nucl. Phys. {\bf A566} (1994) 457c.
\bibitem{levai}		L. V. Bravina et al, Nucl. Phys. {\bf A566} (1994) 461c.
\end{thebibliography}
\end{document}